\documentclass[preprint,proceedings]{rmaa}

\SetYear{2003}
\SetConfTitle{IAUC 194}

\title{INTEGRAL observations of the transient source IGR J19140+098}

\author{
  C. Cabanac\altaffilmark{1}, J. Rodriguez\altaffilmark{3,2}  P.O. Petrucci\altaffilmark{1}, G. Henri\altaffilmark{1}, D. Hannikainen\altaffilmark{4}, P. Durouchoux\altaffilmark{3}
}

\altaffiltext{1}{LAOG, BP 53 F-38041 Grenoble C\'edex 9 France (\protect\email{ccabanac@obs.ujf-grenoble.fr}).}
\altaffiltext{2}{ISDC, Versoix, Switzerland}
\altaffiltext{3}{CEA, Saclay, France}
\altaffiltext{4}{Observatory, University of Helsinki, Finland}

\suppressfulladdresses

\listofauthors{C. Cabanac, J. Rodriguez, P.O Petrucci, G. Henri, D. Hannikainen, P. Durouchoux}
\indexauthor{Cabanac, C., Rodriguez, J., Petrucci, P.O., Henri, G., Hannikainen, D., Durouchoux, P.}

\addkeyword{}

\begin{document}
\maketitle 

\boldabstract{IGR J19140+098 was discovered during the early INTEGRAL observations of GRS1915+105 in March 2003. The following observations by INTEGRAL and RXTE show significant variability on various time scales from 100s to 1ks, but no pulsations. The ISGRI spectra show strong spectral variability in the 20-80keV range. Combined JEMX\_2-ISGRI spectra are well fitted by a power-law and a thermal emission model with a rather high temperature. RXTE data also reveal an apparently broad ionised iron line not detected in earlier INTEGRAL observations. These results are compatible with a galactic X-ray binary source.
}
\section{The Data  and analysis}
IGR J19140+098 (Simbad corrected name IGR J19140+0951) was detected in the FOV of IBIS, the imager aboard INTEGRAL, for the first time during an observation of GRS1915+105 (see \textit{Hannikainen et al. 2003}) on 6th March 2003 ( Spacecraft revolution \#48,  88 ks). The source is about 1 degree close from GRS1915+105 and has been observed after in two other revolutions : Revs. \#69 and \#70 (2003 May 9 and 10-11) for 76 and 96 ks respectively but not in Rev \#59 on 9th April (90 ks ).
RXTE observed the source on March 10th 2003 during 3 ks (public data).
\begin{figure}[!t]
  \includegraphics[width=\columnwidth]{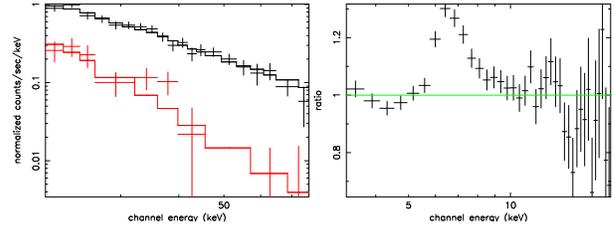}
  \caption{\textit{Left} 20-80 keV ISGRI spectra of the source during rev\#70 separated by 8.6 ks. Photon index are $2.21\pm 0.15$ (black, upper curve) and $3.8\pm 1.0$ (grey, lower curve). \textit{Right} RXTE-PCA data to adopted model ratio (simple powerlaw with $N_H$ fixed to $5.6$ $10^{22}$ $cm^{-2}$ ). }
  \label{fig:}
\end{figure}
The source varies on a monthly timescale and ISGRI light curves show also that the flux can increase by a factor of 3 in 2000s.
The autocorrelation of the RXTE light curves (16s bin size) indicates a typical correlation time of about 150 s. Using the causality argument,  it corresponds to a size of about 3 AU.
No QPO has been identified in the PDS.
The integrated ISGRI flux in the 13-100 keV energy range varied from $1.8$ $10^{-9}erg.s^{-1}.cm^{-2}$ (highest value observed) in rev\#70 to the minimum detection level in rev\# 59.
JEMX\_2 and ISGRI spectra show that the emission can be modeled consistently by a "hard" power law in the 10-100 keV energy range ($\Gamma=1.7$). Adding a "thermic" model for the lower energy emission (multicoulour disc model or simple black-body, freezing the power law parameters) improves the fits (typical temperature obtained : $T_{in}\simeq1.8$ $keV$).
The 20-80keV spectra are well fitted by a simple power law model but show strong spectral variability  (Fig. \ref{fig:}). 
Adding a gaussian line  to JEMX\_2-ISGRI spectra does not reduce the $\chi^2$ significantly whereas it does in RXTE data, performed 4 days later. The line obtained is then consistent with a broad gaussian line ($\sigma = 0.4$ $keV$) centered on 6.7 keV, compatible with an ionised $Fe$ $K_{\alpha}$ line (Fig. \ref{fig:}).
\section{Conclusion and perspectives}
The source IGR J19140+0951 has been discovered by INTEGRAL. The spectra are well-fitted by a power-law + thermal component. A broad and apparently ionised iron line is clearly detected in RXTE public data but not in INTEGRAL observations taken 4 days earlier. The source exhibits strong spectral and flux variabilities more typical of  an X-ray binary than of an AGN. 
 IGR 16320-4751, another X-ray source (re-)discovered by INTEGRAL seem to manifest similar variable behaviour. According to \textit{Rodriguez et al, 2003}, it is most probably a galactic X-ray binaries hosting a neutron star. However we cannot exclude the case of a black-hole binarie concerning IGR J19140+098

\end{document}